\documentclass{article}

\usepackage[final,nonatbib]{neurips_2023_ml4ps}

\usepackage{amsmath,amsfonts,amssymb}
\usepackage[utf8]{inputenc} 
\usepackage[T1]{fontenc}    
\usepackage{hyperref}       
\usepackage{cleveref}
\usepackage{url}            
\usepackage{booktabs}       
\usepackage{nicefrac}       
\usepackage{microtype}      
\usepackage{xcolor}         
\usepackage{graphicx}
\usepackage{glossaries}
\usepackage{subcaption}
\usepackage{tikzducks}
\usepackage{siunitx}
\graphicspath{{figures/}}
\usepackage{placeins}[section]

\usepackage{comment}
\usepackage{todonotes}
\newcommand{\onet}{\hat{w}_\ell^{\text{ONet}}}

\DeclareFontFamily{U}{matha}{\hyphenchar\font45}
\DeclareFontShape{U}{matha}{m}{n}{
      <5> <6> <7> <8> <9> <10> gen * matha
      <10.95> matha10 <12> <14.4> <17.28> <20.74> <24.88> matha12
      }{}
\DeclareSymbolFont{matha}{U}{matha}{m}{n}
\DeclareMathSymbol{\odiv}         {2}{matha}{"63}

\glsdisablehyper
\newacronym{AoA}{AoA}{angle-of-attack}

\newacronym{BC}{BC}{boundary condition}

\newacronym{CFD}{CFD}{computational fluid dynamics}

\newacronym{ONet}{DeepONet}{deep operator network}

\newacronym{FEM}{FEM}{finite element method}
\newacronym{FNO}{FNO}{Fourier neural operator}

\newacronym{GMM}{GMM}{Gaussian mixture model}

\newacronym{ML}{ML}{machine learning}
\newacronym{MLP}{MLP}{multi-layer perceptron}
\newacronym{MSE}{MSE}{mean squared error}
\newacronym{MVE}{MVE}{mean-variance estimation}

\newacronym{NBF}{NBF}{neural basis function}
\newacronym{NN}{NN}{neural network}
\newacronym{NIG}{NIG}{normal inverse-gamma}
\newacronym{NLL}{NLL}{negative log-likelihood}
\newacronym{NSE}{NSE}{Navier-Stokes equations}

\newacronym{OOD}{OOD}{out-of-domain}

\newacronym{PDE}{PDE}{partial differential equation}
\newacronym{PINN}{PINN}{physics-informed neural network}
\newacronym{POD}{POD}{principal orthogonal decomposition}

\newacronym{RAM}{RAM}{Radio Attenuation Measurement}
\newacronym{ROM}{ROM}{reduced order modeling}

\newacronym{SciML}{SciML}{scientific machine learning}
\newacronym{SGD}{SGD}{stochastic gradient descent}
\newacronym{SVD}{SVD}{singular value decomposition}

\newacronym{UQ}{UQ}{uncertainty quantification}

\title{Ensemble models outperform single model uncertainties and predictions for operator-learning of hypersonic flows}
\author{Victor J. Leon$^1$,\,\,\,\,Noah Ford$^2$,\,\,\,\,Honest Mrema$^3$,\,\,\,\,Jeffrey Gilbert$^3$,\,\,\,\,Alexander New$^1$\\
    $^1$ Research and Exploratory Development Department \\
    $^2$ Force Projection Section \\
    $^3$ Air and Missile Defense Sector \\
    Johns Hopkins University Applied Physics Laboratory\\
    Laurel, Maryland 21044\\
    \texttt{\{victor.leon, noah.ford, honest.mrema, jeffrey.gilbert, alex.new\}@jhuapl.edu}
}

\begin{document}

\maketitle

\begin{abstract}
High-fidelity computational simulations and physical experiments of hypersonic flows are resource intensive. Training \gls{SciML} models on limited high-fidelity data offers one approach to rapidly predict behaviors for situations that have not been seen before. However, high-fidelity data is itself in limited quantity to validate all outputs of the \gls{SciML} model in unexplored input space. As such, an uncertainty-aware \gls{SciML} model is desired. The \gls{SciML} model's output uncertainties could then be used to assess the reliability and confidence of the model's predictions. In this study, we extend a \gls{ONet} using three different uncertainty quantification mechanisms: \gls{MVE}, evidential uncertainty, and ensembling. The uncertainty aware \gls{ONet} models are trained and evaluated on the hypersonic flow around a blunt cone object with data generated via computational fluid dynamics over a wide range of Mach numbers and altitudes. We find that ensembling outperforms the other two uncertainty models in terms of minimizing error and calibrating uncertainty in both interpolative and extrapolative regimes.

\end{abstract}

\glsresetall


\section{Introduction}\label{sec:introduction}

Scientists and engineers gain understanding of large, complex systems like weather~\cite{Jordan2017wrf} and flight vehicles~\cite{quinlan2021activelearning} by analyzing databases of how these systems behave, based on input parameters. These instances can be obtained via high-fidelity sources like computational simulation or physical experimentation. However, it is typically infeasible to obtain such data for every parameter configuration of interest. Further data can be generated by \gls{SciML} models that rapidly predict systems behavior for parameters not currently found in databases~\cite{Lu2021deeponet,li2021fno,Witman2022nbf}.

Because the high-fidelity ground truth is limited in quantity, it may not be sufficient to enable training of \gls{SciML} models that can make predictions for the entire parameter space. This motivates the further incorporation of \gls{UQ} into these \gls{SciML} models~\cite{Zhang2019uqpinns,Yang2022uqoperator}. Uncertainties can be used to assess the reliability of predictions, and they can also be used to drive targeted acquisition of further data in an active learning loop~\cite{Arthurs2021activepinn,Li2022multifidelityactive}.

In this paper, we extend the \gls{ONet}~\cite{Lu2021deeponet} using three different \gls{UQ} mechanisms: \gls{MVE}~\cite{Nix1994meanvariance}, evidential uncertainty~\cite{Sensoy2018evidential,Amini2020evidential,Soleimany2021evidential}, and ensembling (\Cref{sec:ml_models}). We evaluate these models on data generated by the steady-state compressible \gls{NSE} with a non-uniform geometry based on a hypersonic flight vehicle (\Cref{sec:data_generation}) in both interpolation and extrapolation settings (\Cref{sec:results}). Although calibration in the extrapolation setting remains challenging to achieve, ensembling consistently outperforms other methods. This echoes findings in fields like chemistry~\cite{Tan2023uqcomparison} and motivates further development of probabilistic operator networks, especially those capable of extrapolating across parameter spaces.

Prior to the development of modern operator networks, \gls{UQ} has been used in engineering fields to accelerate efficient data acquisition. Frequently, techniques like Gaussian process regression are used to predict a single target property like a drag coefficient, based on a small number of input parameters~\cite{quinlan2021activelearning}. Here we consider the more general challenge of predicting pointwise uncertainties associated with a spatially-varying field like velocity. These pointwise uncertainties can still be aggregated into an acquisition function (e.g.,~\cite{Arthurs2021activepinn}), or they can be analyzed in their own right (e.g.,~\cite{Vadeboncouer2023neuralprocesses}).
 

\section{Methods}\label{sec:methods}




\subsection{Problem setup}\label{sec:problem_setup}

We consider a general setting in which our training dataset $\mathcal{D}$ consists of measurement sets $(X, W^d, \psi^d),d=1,\hdots,D$, where $X = \{x_n\}_{n=1}^N\subseteq\Omega\subseteq\mathbb{R}^{N_x}$ is a spatial mesh over a possibly-irregular geometry $\Omega$, shared across all measurement sets, $W^d = \{w^d_n\}_{n=1}^N\subseteq\mathbb{R}^{N_w}$ is a set of state variables values (with $w^d_n$ the value at mesh point $x^d_n$), and $\psi^d\in\mathbb{R}^{N_\psi}$ is a parameter vector. We learn a set of predictive models $\hat{w}_\ell$, one for each state variable $\ell=1,\hdots,N_w$.

\begin{figure}[h]
    \centering
    \includegraphics[width=\linewidth]{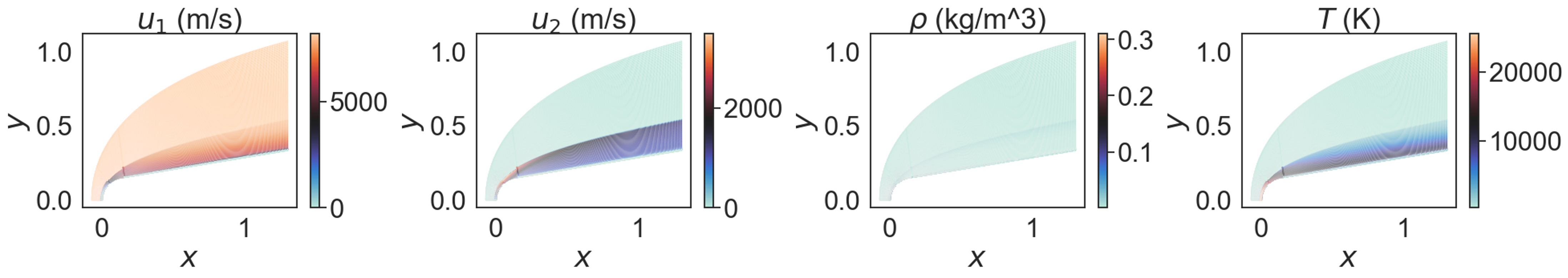}
    \caption{We show the four state variables that are solved for the axisymmetric geometry (\Cref{sec:data_generation}). This solution corresponds to Mach $25$ and altitude \SI{46}{km}.}
    \label{fig:solution}
\end{figure}

In our setting (\Cref{sec:data_generation}), data satisfy a set of compressible \gls{NSE} defined over a 3D axisymmetric geometry based on the \gls{RAM}-C II flight vehicle~\cite{Farbar20213ramcII,Sawicki2021ramcII}. Due to the axisymmetry, there are two spatial degrees of freedom. The state variables are: $x$-velocity ($u_1$), $y$-velocity ($u_2$), density ($\rho$), and temperature ($T$). The parameter vector $\psi$ has two components: Mach number and altitude. \Cref{fig:solution} demonstrates an example solution for this system.

\subsection{Deep operator networks}\label{sec:ml_models}

To solve the RAM-C II fluid flow problem, we rely on the \gls{ONet}~\cite{Lu2021deeponet} model. For a state variable $\ell$, the \gls{ONet} $\hat{w}_\ell$ is a composition of \glspl{MLP}:
\begin{equation}
    (x, \psi) \overset{\hat{w}_\ell}{\mapsto} f_d(f_x(x)\odot f_\psi(\psi)),
    \label{eq:onet}
\end{equation}
where $\odot$ is element-wise multiplication, $f_x$ is an encoder for the spatial points $x$, $f_\psi$ is an encoder for the parameters $\psi$, and $f_d$ is a decoder for the predicted state variable. \glspl{MLP} have $\tanh$ activation functions. Each $\onet$ is trained by using \gls{SGD} and Adam~\cite{Kingma2014adam}, over tuples $(x_n, \psi^d, w^d_n)$ to minimize the squared errors $|w^d_{n,\ell} - \onet(x_n, \psi^d)|^2$. 

We extend the \glspl{ONet} with \gls{MVE}~\cite{Nix1994meanvariance}, evidential~\cite{Amini2020evidential}, and ensemble~\cite{Lakshminarayanan2017ensembles} \gls{UQ} methods. Each enables the \glspl{ONet} to output a mean $\mu_\ell$ and standard deviation $\sigma_\ell$, associated with a specified spatial point $x$ and parameter configuration $\psi$. Precise formulations for each \gls{UQ} approach are standard, but we  give them for reference in~\Cref{sec:uncertainty}. 

\section{Results}\label{sec:results}

We evaluate models with respect to prediction error and uncertainty calibration for both in-domain (interpolating) and \gls{OOD} (extrapolating) settings. Model error is measured with the relative error between a simulation's predicted and ground truth state variables, measured in the normalized state variable space. We also compare the probabilistic \glspl{ONet} to a single (deterministic) \gls{ONet}. Calibration is assessed using miscalibration area, as implemented in the Uncertainty Toolbox~\cite{Chung2021uncertaintytoolbox}. A higher miscalibration area means the model's calibration is worse.

Our dataset (\Cref{sec:data_generation}) solves the system's behavior across variation in the parameter vector $\psi = (\mathrm{Mach}, \mathrm{altitude})$, where $\mathrm{Mach} \in \{10,11,\hdots,30\}$ and $\mathrm{altitude}\in\{20, 22, \hdots,60\}$, in units of kilometers. This yields a total of $441$ simulations. We define the in-domain parameter regime as those simulations having parameters $\psi \in [12, 28]\times[26, 54]$  ($255$ simulations). The \gls{OOD} parameter regime is subdivided into four regions: high Mach ($\psi\in\{29, 30\}\times\{20,22,\hdots,60\}$), low Mach ($\psi \in \{10,11,12\}\times\{20,22,\hdots,60\}$), high altitude ($\psi \in \{10,11,\hdots,30\}\times\{56,58,60\}$, and low altitude ($\psi \in \{10,11,\hdots,30\}\times\{20,22,24\}$). For evaluating in-domain interpolation and comparing it to out-of-domain extrapolation, we sample $50$ simulations from the in-domain regime to use as a test set. All models use the same hyperparameters and training settings (\Cref{sec:hyperparameters}).

\begin{figure}
    \centering
    \includegraphics[width=0.7\linewidth]{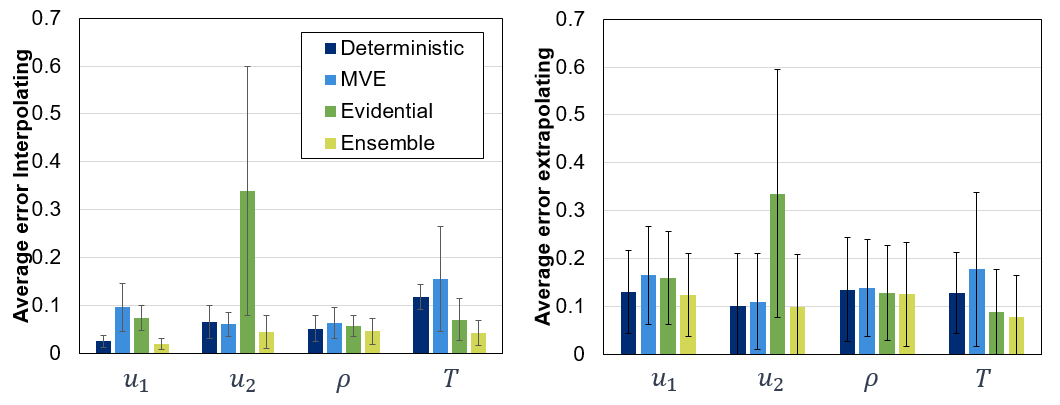}
    \caption{Average relative errors interpolating (in-domain) and extrapolating (\gls{OOD}) for the four model types across state variables. When interpolating, the ensemble model has the lowest error, with the deterministic model a close second. When extrapolating, the four models perform more similarly, although on average, the ensemble model has the lowest error. For both interpolation and extrapolation, the \gls{MVE} and evidential models have worse spread and error when predicting $T$ and $u_2$, respectively. The bars indicate average error across the domain's test set for a given parameter. The error bars are the inter-quartile range of the errors across the domain's test set. The ensemble model performs best when averaging over state variables in all domains, too (\Cref{tab:results} in~\Cref{sec:extra_stuff}).
    }
    \label{fig:errors}
\end{figure}

\begin{figure}
    \centering
    \includegraphics[width=0.7\linewidth]{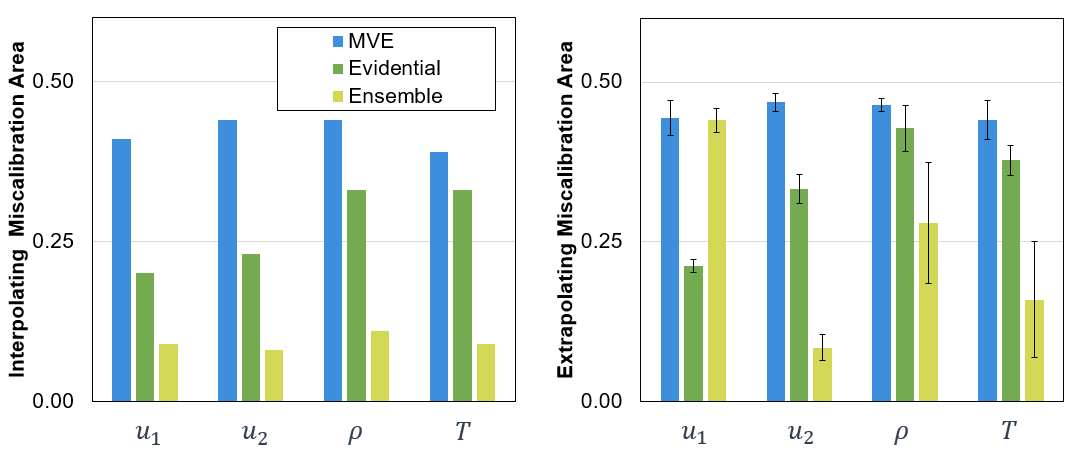}
    \caption{Uncertainty miscalibration areas interpolating (in-domain) and extrapolating (\gls{OOD}) for the three probabilistic model types across state variables. When interpolating, the ensemble model has the lowest miscalibration area, which means that the uncertainties are well calibrated. The \gls{MVE} and evidential models have significantly higher miscalbration areas. When extrapolating, the ensemble model has the lowest miscalibration area. No model has good uncertainty calibration when extrapolating for $u_1$. Interestingly, the evidential model has significantly lower miscalibration area than the \gls{MVE} model for all state variables in-domain and for $u_1$ \gls{OOD} than the other two models, implying that the evidential model has better uncertainty calibration. In fact, for some spatial points, the evidential model predicts large uncertainties of order $10^4$ for $u_1$ and $u_2$ and $10^1$ for $\rho$ and $T$ (\Cref{fig:miscalibration_evidential} in~\Cref{sec:extra_stuff}), meaning that the evidential model is not converging on calibrated uncertainties. This is not captured by miscalibration area since it is insensitive to small numbers of outlier uncertainties. These uncertainties are far too large, as all state variables are normalized by mean and standard deviation of the train set before training, and thus, are all order \textasciitilde 1. The bars indicate the miscalibration area across the test domain. The error bars in extrapolation are the standard deviation of miscalibration area across the four extrapolation regimes (high Mach, high altitude, low Mach, low altitude).
    }
    \label{fig:miscalibration}
\end{figure}

In~\Cref{fig:errors}, we show that, in-domain, the ensemble model has the lowest errors for all state variables, with the deterministic model a close second. The \gls{MVE} and evidential model have significantly higher errors across all state variables. Out-of-domain, all models perform more similarly, but on average, the ensemble model still has lower error. Turning to uncertainty calibration in-domain, the ensemble model has by far the best calibration (\Cref{fig:miscalibration}). As training epochs increase, we observe that evidential model does not converge to meaningful uncertainty magnitudes. Out-of-domain, the ensemble model again has the best calibration of the three models. 

\begin{figure}
    \centering
    \includegraphics[width=\linewidth]{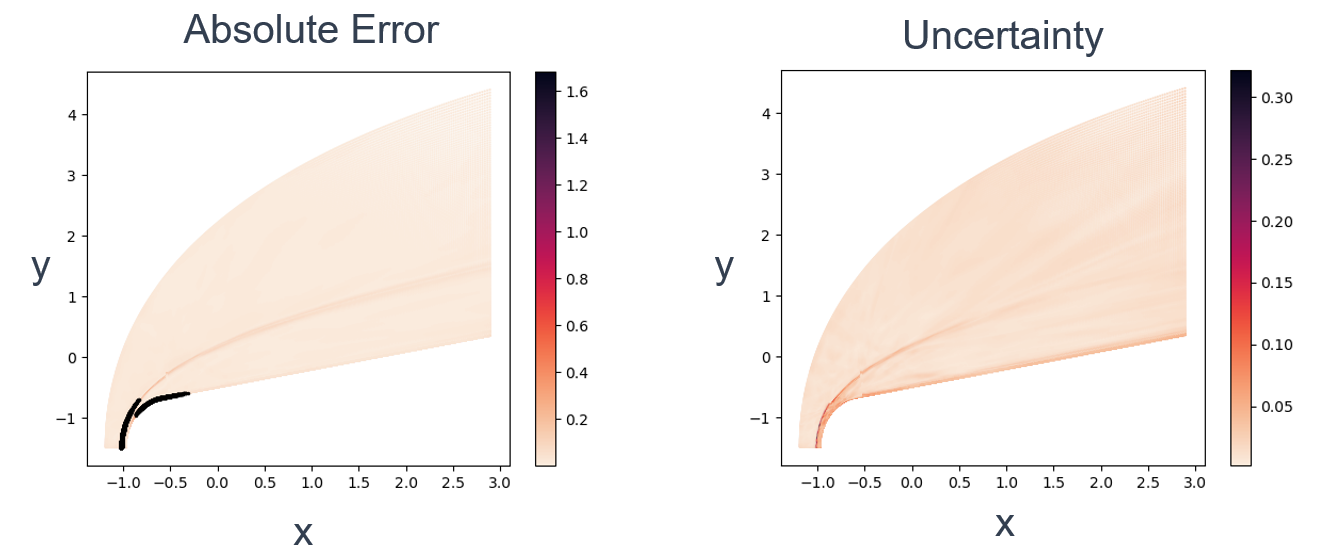}
    \caption{Spatial regions of higher uncertainty are correlated to regions of higher pointwise, absolute error in prediction for $u_1$. Darker points in both plots correspond to higher magnitudes of error and uncertainty. In the error plot, regions of relatively high error ($>0.5$) are highlighted in black. Higher error and uncertainty regions are concentrated where state variables rapidly change in value over short distances and at discontinuities (i.e. at the bow shock and at the surface boundary). This plot is from the in-domain test set, with Mach number $25$ and altitude \SI{46}{km}. Similar trends were observed across state variables and parameters. 
    }
    \label{fig:spatial_U}
\end{figure}

None of the models have good calibration for $u_1$ when extrapolating. To understand why, we plot prediction error and uncertainty spatially (\Cref{fig:spatial_U}), which reveals that uncertainty and error are spatially correlated. Regions of higher error and uncertainty are in regions in which $u_1$ changes rapidly over small spatial distances (i.e. at the hypersonic bow shock, in the boundary layer at the front tip of the cone, and at the surface boundary of the blunt cone). This implies that the \gls{ONet} has difficulty fitting rapid spatial changes in state variables.

\section{Conclusion}\label{sec:conclusion}

We extend the \gls{ONet} with three different \gls{UQ} mechanisms: \gls{MVE}, evidential uncertainty, and ensembling. We observe that the ensemble model outperforms both other UQ mechanisms in the in-domain (interpolative) and \gls{OOD} (extrapolative) regimes for a case study on predicting $u_1$, $u_2$, $\rho$, and $T$ in the hypersonic flow around a blunt nose cone at various Mach numbers and altitudes. For the ensemble model, higher uncertainty is spatially correlated with higher error, which both tend to be concentrated in regions of large changes in state variable values over small distances. This motivates further research into the use of models that inherently account for nonlocal phenomena, such as \glspl{NBF} \cite{Witman2022nbf}, POD-DeepONets \cite{Lu2022poddeeponet}, and \glspl{FNO} \cite{li2021fno}.

\section*{Acknowledgments}

This work was supported by internal research and development funding from the Air and Missile Defense Sector of the Johns Hopkins University Applied Physics Laboratory.

\bibliographystyle{unsrt}
\bibliography{references}
\clearpage

\appendix

\section{Supplemental figures and tables}\label{sec:extra_stuff}

\begin{figure}[h]
    \centering
    \includegraphics[width=0.4\linewidth]{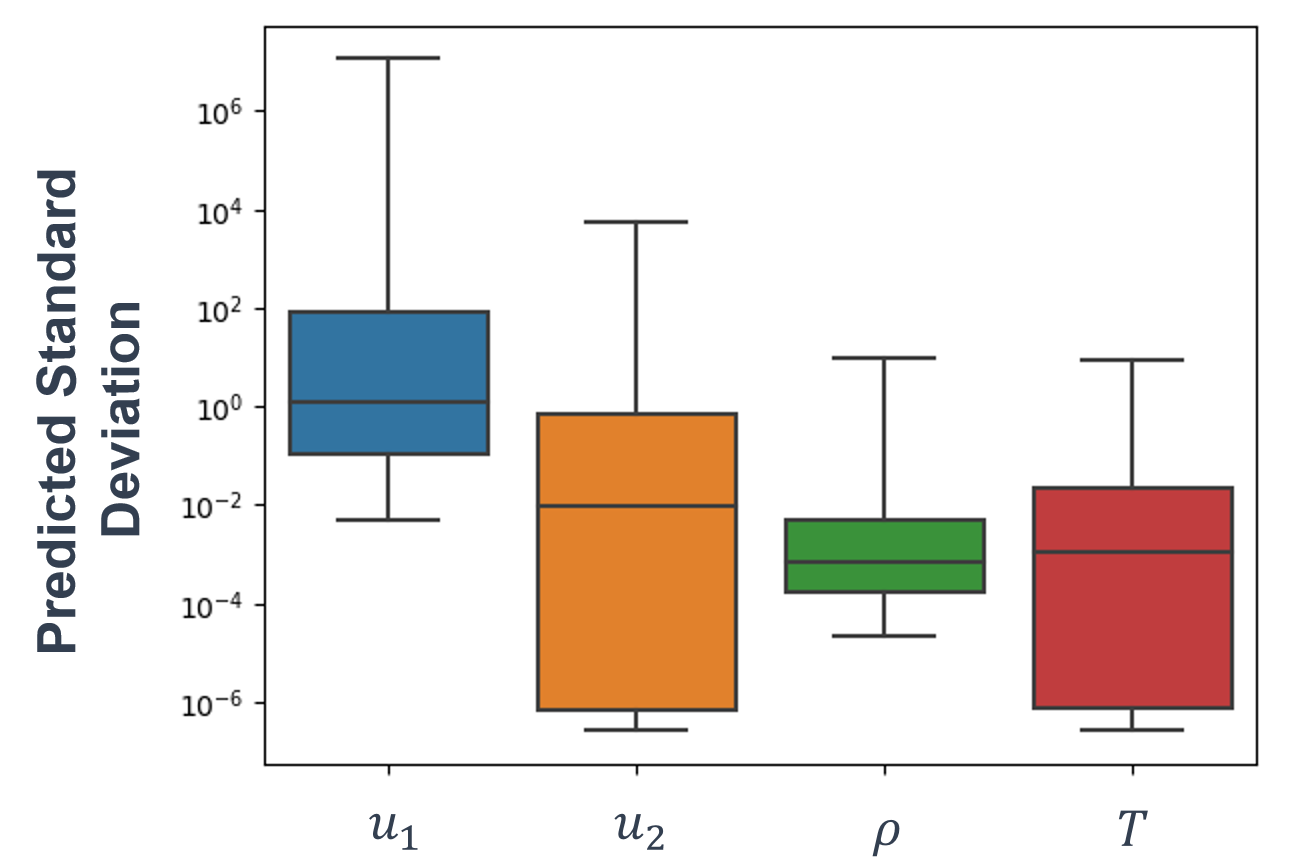}
    \caption{The evidential model does not converge on calibrated uncertainties. All data is normalized by mean and standard deviation, so predicted values are of magnitude \textasciitilde 1. The evidential model predicts uncertainties greater than $10^4$ (underconfident) for $u_1$ and $u_2$ and order $10^1$ for $\rho$ and $T$. This is not captured by miscalibration area since the metric is insensitive to small numbers of outlier uncertainties. }
    \label{fig:miscalibration_evidential}
\end{figure}

\begin{table}[ht]
    \centering
    \begin{tabular}{c|c||c|c}
    UQ mechanism  & Evaluation domain & Mean Absolute Error & Calibration \\\hline\hline
    Deterministic & In-domain         &  $0.0649 \pm 0.0340$        & n/a         \\
                  & High Mach         &  $0.140 \pm 0.0175$        & n/a         \\
                  & High Altitude     &  $0.0996 \pm 0.0135$        & n/a         \\
                  & Low Mach          &  $0.130 \pm 0.0174$        & n/a         \\
                  & Low Altitude      &  $0.124 \pm 0.0356$        & n/a         \\\hline
    MVE           & In-domain         &  $0.0939 \pm .0378$        & $0.42 \pm 0.02$            \\
                  & High Mach         &  $0.173 \pm 0.0436$        & $0.45 \pm 0.03$            \\
                  & High Altitude     &  $0.136 \pm 0.0351$        & $0.45 \pm 0.00$            \\
                  & Low Mach          &  $0.143 \pm 0.0219$        & $0.48 \pm 0.01$            \\
                  & Low Altitude      &  $0.139 \pm 0.0367$        & $0.44 \pm 0.03$            \\\hline
    Evidential    & In-domain         &  $0.135 \pm 0.118$        &  $0.27 \pm 0.06$           \\
                  & High Mach         &  $0.245 \pm 0.200$        &  $0.32 \pm 0.07$           \\
                  & High Altitude     &  $0.158 \pm 0.0871$        & $0.35 \pm 0.10$            \\
                  & Low Mach          &  $0.141 \pm 0.0253$        & $0.35 \pm 0.08$            \\
                  & Low Altitude      &  $0.168 \pm 0.0832$        & $0.34 \pm 0.08$            \\\hline
    Ensembling    & In-domain         &  \boldmath$0.0378 \pm 0.0105$        &$0.09 \pm 0.01$             \\
                  & High Mach         &  \boldmath$0.119 \pm 0.0200$        & $0.24 \pm 0.14$            \\
                  & High Altitude     &  \boldmath$0.0846 \pm 0.0185$        &$0.22 \pm 0.15$             \\
                  & Low Mach          &  \boldmath$0.118 \pm 0.0213$        & $0.27 \pm 0.12$            \\
                  & Low Altitude      &  \boldmath$0.105 \pm 0.0418$        & $0.24 \pm 0.18$           
    \end{tabular}
    \caption{We evaluate the four model types on in-domain and \gls{OOD} prediction tasks in terms of accuracy and calibration. In contrast to \Cref{fig:errors}, which averages over all domains for each state variable, here we report mean absolute errors and standard deviations calculated over all points in each domain's test set averaged over the four state variables. Consistent with \Cref{fig:errors}, we observe that ensembling has the lowest mean absolute errors for all evaluated models (in bold text) and, on average, the best calibration.}
    \label{tab:results}
\end{table}

\section{Data generation}\label{sec:data_generation}
As training and evaluation data, we simulate high Mach number flow over a blunt nose cone, for a set of state variables governed by the compressible \gls{NSE}. The geometry for our study is based on the second flight test from the \gls{RAM} flight experiments performed in the 1960s: the RAM-C II vehicle~\cite{Farbar20213ramcII,Sawicki2021ramcII}. We represent the RAM-C II vehicle by an axisymmetric spherical blunt nose cone. The nose radius is \SI{0.1524}{m} and connects tangentially to the cone body, which has a half-cone angle of \SI{9}{^\circ}. The full body length of the configuration is \SI{1.3}{m}. We use CFD++ version 20.1~\cite{web_cfd++} to generate ground truth simulations.


\section{Network hyperparameters}\label{sec:hyperparameters}

\Cref{tab:onet_params} gives the hyperparameters used to train the \glspl{ONet} in this study. We do not find that evidential model performance varies significantly with evidential regularization hyperparameter $\lambda$ for values in the range  $[10^{-3}, 1]$.

\begin{table}[h]
    \centering
    \begin{tabular}{c|c}
    Hyperparameter      &   Value\\\hline\hline 
    \# of hidden units for the spatial encoder $f_x$        &    $32$\\
    \# of layers for the spatial encoder $f_x$              &    $1$\\
    \# of hidden units for the parameter encode $f_\psi$    &    $32$\\
    \# of layers for the parameter encoder $f_\psi$         &    $1$\\
    \# of hidden units for the decoder $f_d$                &    $256$\\
    \# of layers for the decoder $f_d$                      &    $3$\\
    weight decay for training the \glspl{ONet}              &    $10^{-4}$\\
    \# epochs for training the \glspl{ONet}                 &    $97$\\
    evidential regularization strength $\lambda$            &    $10^{-2}$\\
    \# of ensemble members                                  &   $10$
    \end{tabular}
    \caption{Hyperparameters for training the \gls{ONet} models (Eq.~\ref{eq:onet})}
    \label{tab:onet_params}
\end{table}

\section{Uncertainty quantification for operator-learning}\label{sec:uncertainty}

We evaluate three schemes for \gls{UQ} in this work. \gls{MVE} (\Cref{sec:mean_variance}) and evidential uncertainty (\Cref{sec:evidential}) are probabilistic methods that extend the network architecture and loss function, while ensembling (\Cref{sec:ensembling}) is not. Other techniques not considered here include dropout~\cite{Gal20216dropout} and \glspl{GMM}~\cite{Zhu2023gmm}.

\subsection{Mean-variance estimation}\label{sec:mean_variance}

In \gls{MVE}, state variables are assumed to follow a conditionally normal distribution. The \gls{ONet} $\hat{w}^\ell$ for the state variable $\ell$ has two output variables, the mean $\mu_\ell$ and variance $\sigma_\ell^2$:

\begin{eqnarray}
    (x, \psi) &\overset{\hat{w}_\ell}{\mapsto}& \mu_\ell(x, \psi), \sigma^2_\ell(x, \psi),\\
    \mu_\ell(x, \psi) &=& [f_d(f_x(x)\odot f_\psi(\psi))]_1 \nonumber\\
    \sigma^2_\ell(x, \psi) &=& \mathrm{SoftPlus}[f_d(f_x(x)\odot f_\psi(\psi))]_2 \nonumber,
    \label{eq:mve_pred}
\end{eqnarray}
where the $\mathrm{SoftPlus}$ ensures nonnegativity of variance. \gls{MVE} \glspl{ONet} are trained by minimizing the Gaussian \gls{NLL}:

\begin{equation}
    \mathrm{Loss}((x, \psi), w_\ell) = \log \sigma_\ell^2(x, \psi) + \frac{(\mu_\ell(x, \psi) - w_\ell)^2}{\sigma_\ell^2(x, \psi)},
    \label{eq:mve_loss}
\end{equation}
for a partial data point $((x, \psi), w_\ell) \in \mathbb{R}^{N_x}\times\mathbb{R}^{N_\psi}\times\mathbb{R}$.

\subsection{Evidential uncertainty}\label{sec:evidential}

In evidential uncertainty~\cite{Sensoy2018evidential,Amini2020evidential,Soleimany2021evidential}, a \gls{NIG} probabilistic model is imposed on the data:

\begin{eqnarray}
    w_\ell | x, \psi &\sim& \mathcal{N}(\mu_\ell, \sigma^2_\ell) \\
    \mu_\ell | x, \psi &\sim& \mathcal{N}(\gamma_\ell, \sigma^2_\ell / v_\ell)\nonumber\\
    \sigma^2_\ell | x, \psi &\sim& \Gamma^{-1}(\alpha_\ell, \beta_\ell).\nonumber
\end{eqnarray}

From this hierarchical model, we have the predicted state variable value as $\mathbb{E}[\mu_\ell] = \gamma_\ell$, the predicted aleatoric uncertainty as $\mathbb{E}[\sigma^2_\ell] = \beta_\ell / (\alpha_\ell - 1)$, and the predicted epistemic uncertainty as $\mathrm{Var}[w_\ell] = \beta_\ell / (v_\ell (\alpha_\ell - 1))$. Thus, the predictive model $\hat{w}_\ell$ for state variable $\ell$ is given by:
\begin{eqnarray}
    x,\psi &\overset{\hat{w}_\ell}{\mapsto}& \gamma_\ell,v_\ell,\alpha_\ell,\beta_\ell\label{eq:prediction}\\
    \gamma_\ell &=& [f_d(f_x(x)\odot f_\psi(\psi))]_1\nonumber\\
    v_\ell &=& \mathrm{SoftPlus}[f_d(f_x(x)\odot f_\psi(\psi))]_2\nonumber\\
    \alpha_\ell &=& \mathrm{SoftPlus}[f_d(f_x(x)\odot f_\psi(\psi))]_3 + 1\nonumber\\
    \beta_\ell &=& \mathrm{SoftPlus}[f_d(f_x(x)\odot f_\psi(\psi))]_4,\nonumber
\end{eqnarray}
where each $f_\ell$ is a \gls{NN}.

For a partial datapoint $((x,\psi),w_\ell)\in\mathbb{R}^{N_X}\times\mathbb{R}^{N_W}\times\mathbb{R}$, the evidential loss function~\cite{Amini2020evidential} is:
\begin{eqnarray}
    \mathrm{Loss}((x,\psi), w) &=& \sum_\ell \mathrm{Loss}^{NLL}((x,\psi), w_\ell) + \lambda\,\mathrm{Loss}^R((x,\psi), w_\ell)
    \label{eq:evidential_loss},
\end{eqnarray}
where $\mathrm{Loss}^{NLL}$ is a \gls{NLL}:
\begin{eqnarray}
    \mathrm{Loss}^{NLL}((x,\psi), w_\ell) &=& \frac{1}{2}\log\frac{\pi}{v_\ell}-\alpha_\ell \log\Omega_\ell + \log\frac{\Gamma(\alpha_\ell)}{\Gamma(\alpha_\ell + (1/2))}\nonumber\\
    && + \left(\alpha_\ell + \frac{1}{2}\right) \log\left((w_\ell - \gamma_\ell^2 v_\ell + \Omega_\ell\right)\\
    \Omega_\ell &=& 2\beta_\ell(1 + v_\ell)\nonumber
\end{eqnarray}
and $\mathrm{Loss}^R$ is a regularization term:
\begin{eqnarray}
    \mathrm{Loss}^R((x,\psi), w_\ell) &=& |w_\ell - \gamma_\ell| (2v_\ell + \alpha_\ell)
\end{eqnarray}

\subsection{Ensembling}\label{sec:ensembling}

In ensembling~\cite{Lakshminarayanan2017ensembles}, a set of $B$ models $\hat{w}_{\ell,1},\hdots,\hat{w}_{\ell,B}$ are independently trained from different network weight initializations. Then the predicted means and standard deviations are $\mu_\ell(x,\psi) = \underset{b}{\mathrm{mean}}\{ \hat{w}_{\ell,b}\}$ and $\sigma_\ell(x, \psi) = \underset{b}{\mathrm{StdDev}}\{\hat{w}_{\ell, b}\}$.

\end{document}